\documentclass[runningheads]{llncs}
\usepackage{graphicx}
\usepackage{subfigure}
\usepackage{lineno,hyperref}
\modulolinenumbers[5]
\usepackage{amssymb}
\usepackage{latexsym}
\usepackage{diagbox}
\usepackage{amsmath}
\usepackage{amssymb}
\usepackage{xcolor}
\usepackage{overpic}
\usepackage{mathtools}
\usepackage{nth}
\usepackage{amsmath}
\usepackage{multirow}
\usepackage{dsfont}
\usepackage{float}
\usepackage{booktabs}
\usepackage{xcolor}
\usepackage{graphicx}
\usepackage{bbm}
\usepackage{adjustbox}
\usepackage[misc,geometry]{ifsym}
\graphicspath{{figures/}}
\makeatletter
\def\thanks#1{\protected@xdef\@thanks{\@thanks
        \protect\footnotetext{#1}}}
\makeatother

\begin{document}
\title{Domain Generalization for Mammography Detection via Multi-style and Multi-view Contrastive Learning}
\titlerunning{Domain Generalization for Mammography Detection}
%
\author{Zheren Li\inst{1,2}\orcidID{0000-0003-3787-3940} \and
Zhiming Cui\inst{1,3} \and
Sheng Wang\inst{1,2} \and
Yuji Qi\inst{4} \and 
Xi Ouyang\inst{1,2} \and
Qitian Chen\inst{1} \and
Yuezhi Yang\inst{3} \and
Zhong Xue\inst{1} \and
Dinggang Shen\inst{1,5} \and
Jie-Zhi Cheng\inst{1}\Letter
\thanks{Z. Li and Z. Cui---Equal contribution.}
}
\authorrunning{Z. Li et al.}

\institute{Shanghai United Imaging Intelligence Co., Ltd., China \\
\email{jzcheng@ntu.edu.tw} \and
Institute for Medical Imaging Technology, School of Biomedical Engineering, Shanghai Jiao Tong University, Shanghai, China \and 
Department of Computer Science, The University of Hong Kong, Hong Kong \and
Department of Biomedical Engineering, Yale University, New Haven, USA 
\and School of BME, ShanghaiTech University, Shanghai, China
}

\maketitle         
\begin{abstract}

Lesion detection is a fundamental problem in the computer-aided diagnosis scheme for mammography. The advance of deep learning techniques have made a remarkable progress for this task, provided that the training data are large and sufficiently diverse in terms of image style and quality. In particular, the diversity of image style may be majorly attributed to the vendor factor. However, the collection of mammograms from vendors as many as possible is very expensive and sometimes impractical for laboratory-scale studies. Accordingly, to further augment the generalization capability of deep learning model to various vendors with limited resources, a new contrastive learning scheme is developed. Specifically, the backbone network is firstly trained with a multi-style and multi-view unsupervised self-learning scheme for the embedding of invariant features to various vendor-styles. Afterward, the backbone network is then recalibrated to the downstream task of lesion detection with the specific supervised learning. The proposed method is evaluated with mammograms from four vendors and one unseen public dataset. The experimental results suggest that our approach can effectively improve detection performance on both seen and unseen domains, and outperforms many state-of-the-art (SOTA) generalization methods.

\end{abstract}

\section{Introduction} 

\begin{figure}[t]
\label{fig:different-style}
\centering
\includegraphics[width=0.99\textwidth]{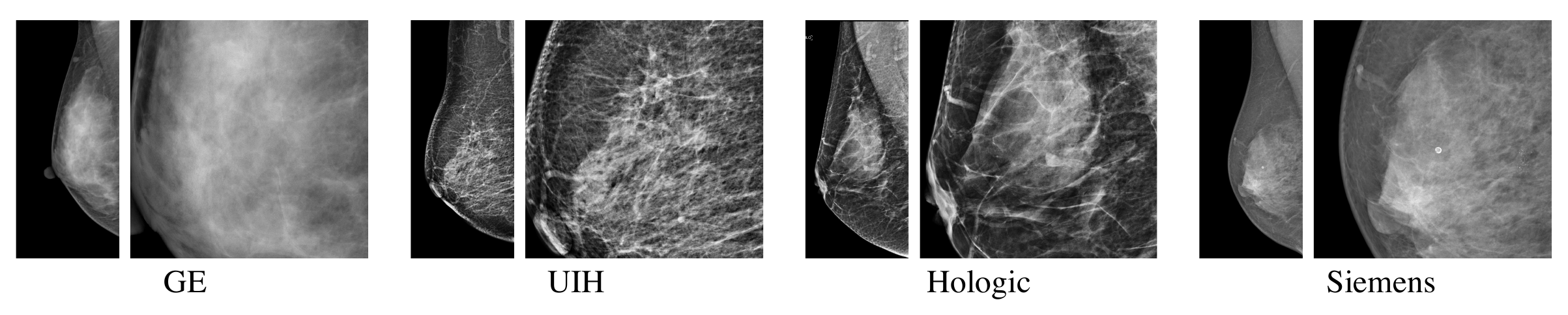}
\caption{Style differences among four mammography vendors, i.e., GE, United Imaging Healthcare(UIH), Hologic, and Siemens. All images of the four vendors are MLO views. To illustrate the detailed image variation, a zoom-in image is provided on the right of each MLO view.}
\end{figure}

The advance of deep learning (DL) techniques have remarkably improved the computer-aided detection (CADe) of breast lesions in mammography. It has been shown in many studies \cite{lotter2021robust,mckinney2020international,salim2020external} that the incorporation of DL-based CADe software in the reading workflow of mammography can effectively improve the detection accuracy. Promising DL-based CADe performance requires large and diverse training data. In particular, the inclusion of wide variety of vendors is very important to equip the DL-based CADe system with prominent generalization capability. As shown in Fig. \ref{fig:different-style}, the styles of images from various vendors vary significantly.
Accordingly, the generalization of DL-based CADe system may be limited if the data of each vendor are not sufficiently included in the training stage. However, the collection of large and diverse data with various vendors is very expensive and is impractical. Meanwhile, it is also well-known that there exists domain gap between datasets from various institutes. Therefore, a domain generalization method is needed to alleviate the demands of large and diverse data from various vendors for the DL-based CADe scheme.


In the literature, the domain generalization for DL technique can be classified into three categories: 1) conventional data argumentation methods, e.g. rotation, flip, deformation and jittering \cite{zhang2020generalizing,romera2018train,volpi2018generalizing}, 2) learning based methods with the generative deep neural networks \cite{yue2019domain,wang2020mr,kim2019diversify,zakharov2019deceptionnet}, 3) learning based methods for the exploration of task-specific and domain-invariant features \cite{wang2020learning,liu2020shape,dou2019domain,li2018domain,li2018deep,cui2021structure,chen2020unsupervised}. 
However, all these learning based methods relied on annotated data, which are costly obtain and availability is relatively limited. A new method to automatically mine domain-invariant features from large unannotated data is very needed.

In this study, to address the above mentioned issues, we explore the contrastive learning technique to augment the generalization capability of DL-based CADe in self-supervised fashion. The contrastive learning has also been shown to generate better pretrained models for several medical image problems, e.g., diagnosis of chest radiography \cite{azizi2021big,sowrirajan2020moco,zhou2020comparing}, dermatology images \cite{azizi2021big}, etc., but less exploited for extracting domain invariant features.
To specifically address the issue of vendor domain gap, we propose a multi-style and multi-view contrastive learning method to boost the lesion detection performance in mammography. 
Specifically, to attain the goal of generalization robustness to multiple vendor styles, the CycleGAN \cite{zhu2017unpaired} technique is employed to generate multiple vendor-style images from a single vendor style image. The generated multiple style images from the same source are randomly paired as positive samples for the multi-style contrastive learning. Meanwhile, for the multi-view contrastive learning, the CC and MLO views of the same breast are also paired as positive samples. After self-supervised training, the backbone of the contrastive learning model is employed for the downstream detection task.

To our best knowledge, this is the first study that explores the contrastive learning to explicitly address the issue of vendor-style domain gap. We specifically train our method on three vendor domains, and then test on images from both seen and unseen domains. The results demonstrated that our method can effectively boost the lesion detection performance on either seen or unseen domain. We also compare our multi-style and multi-view contrastive learning to several state-of-the-art domain generalization methods and suggest our pretrained model can yield better generalization effect on both seen and unseen data.  

\section{Method} 

Fig.\ref{fig:msvcl} illustrates our domain generalization framework for lesion detection in mammography. We first conduct multi-style and multi-view contrastive learning scheme to extract domain invariant features on the data from different vendors. 
Afterward, the pretrained network backbone is transferred to the downstream task of lesion detection, especially on the data from unseen vendors.

\subsection{Contrastive Learning Scheme}
Contrastive learning is a self-supervised learning method, which trains an encoder to quantify the image representation into a proper vector space without the supervision of annotations. The derived model from contrastive learning can be served as a pretrained model for various downstream tasks like segmentation, detection, etc. The major advantage may lie in the flexibility to the downstream tasks.

The basic idea of contrastive learning is to bundle diversified images of the same class/object/subject as positive pairs for the exploration of reliable representation in feature space by the self-supervised learning manner. Specifically, given a  mini-batch of $N$ images, each example is randomly augmented twice with the diversifying operations, e.g., cropping, rotation, style-transferring, etc., to generate an augmented mini-batch with $2N$ samples. In the augmented mini-batch, two samples from the same image source are treated as a positive pair $(i,j)$, whereas the other $2(N-1)$ samples within the mini-batch are regarded as negative pairs. With the positive and negative pairs, the contrastive learning is driven with the contrastive loss to maximize the agreement for the positive pairs. The contrastive loss is defined as:

\begin{equation}
\ell_{i,j} = -\log \frac{{\rm exp}({\rm sim}(z_i,z_j)/\tau)}{\begin{matrix} \sum_{k=1}^{2N} \mathbbm{1}_{[k\neq i]}{\rm exp}({\rm sim}(z_i,z_k)/\tau) \end{matrix}},
\label{eq:contrastive_loss}
\end{equation}
where ${\rm sim}(\cdot)$ is the dot product and $z$ refers to the extracted features. $\mathbbm{1}_{[k\neq i]} \in \{0,1\}$ is an indicator function equaling 1 when $k\neq i$ and $\tau$ is a temperature parameter. 

Via maximizing the agreement for the positive pairs, the learnt features are supposed to be more general and robust to the diversifying operations. We further leverage the concept of contrastive learning to explore the generalization for various vendor styles and the domains of CC and MLO views. The details of multi-style and multi-view contrastive learning are elaborated in the following section.

\subsection{Multi-Style and Multi-View Contrastive Learning}  

\begin{figure}[!t]
  \centering
  \includegraphics[width=1\textwidth]{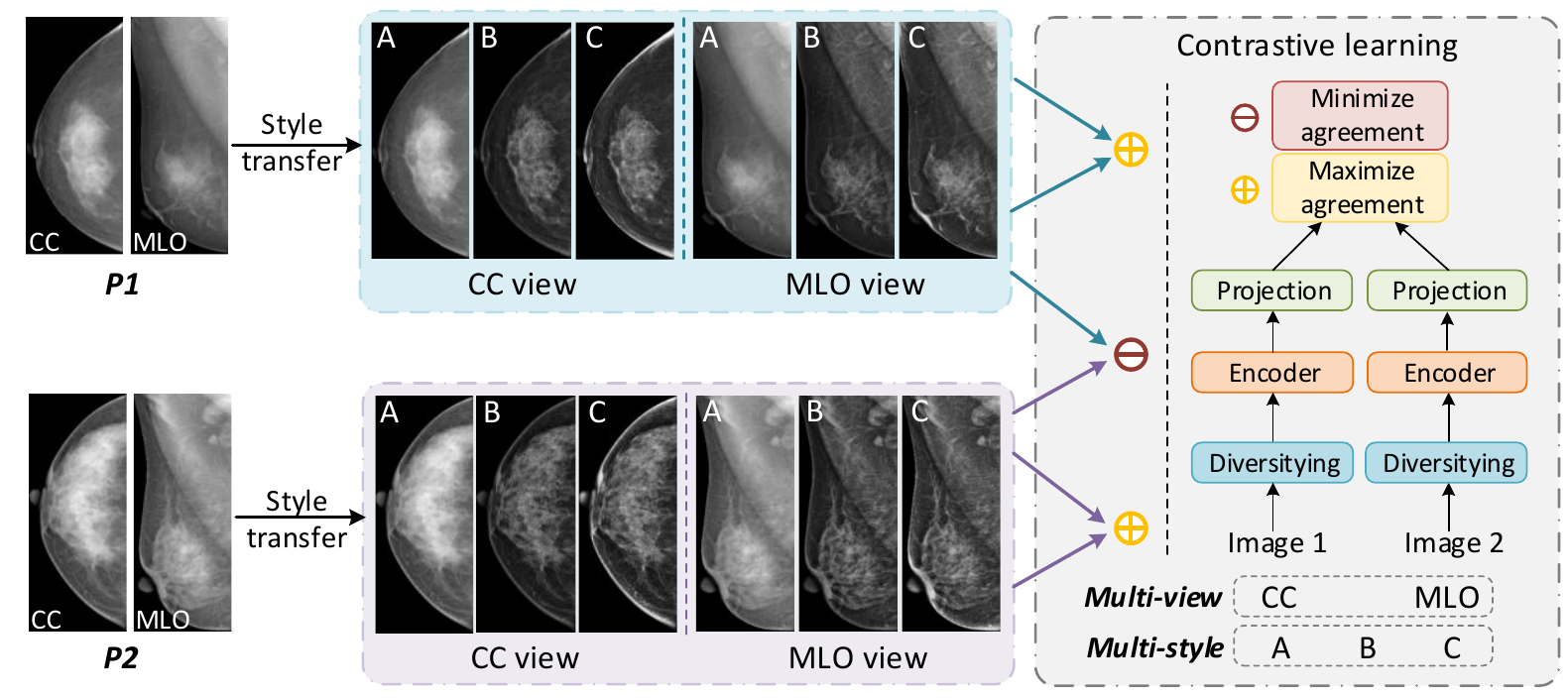}
  \caption{Illustration of multi-style and multi-view contrastive learning.}
  \label{fig:msvcl} 
\end{figure}


\subsubsection{Multi-Style Contrastive Learning} 
Referring to Fig. \ref{fig:msvcl}, the image styles vary to different vendors and are regarded as distinctive domains. To endow the backbones of the lesion detectors better vendor-style generalization capability, we exploit the CycleGAN \cite{zhu2017unpaired} technique as the diversifying operation for the contrastive learning. Specifically, given $M$ seen vendor-style domain, we train $\tbinom{M}{2}$ generators, which map the data distribution of source domain $\Omega_i$ to target $\Omega_j$ domain, $\forall i, j \in M$.
Comparing to the method used in \cite{wang2020mr}, the CycleGAN realizes the style transfer with bidirectional learning process. The work \cite{wang2020mr} unidirectionally takes a few references images may attain limited transferring effect.

With the $\tbinom{M}{2}$ generators, each of the original $N$ images in the training set can be diversified into $M-1$ transferred images. Afterward, a mini-batch with $M \times N$ images can be formed. The positive pairs for the contrastive learning are constituted with any two images of the same image source in the original $N$ image set. Therefore, there are possible $N \times \tbinom{M}{2}$ positive pairs that be randomly selected in the contrastive learning, whereas the other combination pairs are treated as negative pairs. The multi-style contrastive learning is then carried with minimizing the equation \ref{eq:contrastive_loss} to seek feature embedding space with better generalization capability to various vendor domains.


\subsubsection{Multi-View Contrastive Learning}
A standard examination of mammography consists of two CC and MLO views. Because the two standard views are mutually complementary, the appearance of CC and MLO images are different. 
For exampple, a MLO view includes axilla region, while a CC view doesn't. Accordingly, domain gap between these two views may exist. To seek view domain-invariant feature embedding for better lesion detection performance, 
we explore contrastive learning scheme to consider the distinctive view domains. Specifically, we treat the CC and MLO view of the same breast from the same patient as positive pair, whereas the other combination of CC and MLO is a negative pair. To further enrich the sample diversity, we further implement diversifying operations including random cropping, random rotation, horizontal flipping, and adjustment of brightness, contrast, and saturation for the CC and MLO in each positive or negative pair. With the prepared sample pairs, the multi-style contrastive learning can be carried out for the embedding of view-invariant features.


The synergy of the multi-style and multi-view contrastive learning is realized as illustrated in Fig. \ref{fig:msvcl}. Specifically, we firstly employ the style transferring CycleGAN to derive diversified vendor styles and perform the random diversifying operations on the CC and MLO views as the sample pairs for contrastive learning. We will show that our multi-style and multi-view contrastive learning can boost the lesion detection tasks effectively on either seen and unseen domains.

\subsection{Lesion detection Detection with Contrastive Learning}


In this study, we employ the classic detection network of FCOS \cite{tian2019fcos} to identify mass and clustered calcifications in mammography. The derived pretrained model from the contrastive learning is adopted as the backbone of the FCOS architecture for the realization of the downstream lesion detection tasks.
\section{Experiments}
\subsubsection{Implementation Details}
Our method was developed based on Pytorch package and trained with NIVIDA Titan RTX GPUs. We train CycleGAN \cite{zhu2017unpaired,zhang2018task} models with the settings of 50, 100, 200 epochs, and choose the setting, i.e., 100 epochs, that can achieve best lesion detection performance on the validation set. The backbone of the generator is ResNet with 9 blocks with 20 conv layers, while the discriminator is PatchGAN with 6 conv layers for binary classifications. In training, the sizes of inputs and outputs are 512*512 by random cropping the images. MSE and L1 losses are used for classification and reconstruction. For a style transfer, e.g., A to B, the involved training data from A and B are both 1000 for balanced training. Code and models of style transfer are available at: 
https://github.com/lizheren/MSVCL\_MICCAI2021.

We adopt ResNet-50 as the backbone model for contrastive learning and FCOS detector. For fair comparison, the learning rate and batch size for all contrastive learning schemes are set the same as 0.3 and 256, respectively. Meanwhile, all contrastive learning schemes in all experiments use the same diversifying operations, including random cropping, random rotation in $\pm10^{\circ}$, horizontal flipping and random color jittering (strength$=$0.2).
For the training of FCOS models, the SGD method is adopted with the parameters of learning rate, weight decay and momentum set as $0.005$, $10^4$, $0.9$, respectively. The epoch and batch size are set to 50 and 8, respectively, throughout all experiments. Several augmentation methods, e.g. random flipping, scaling, etc., are also implemented in the training of FCOS.

\subsection{Datasets}
In this paper, mammograms of four vendors, i.e., GE, United Imaging Healthcare(UIH), Hologic, and Siemens, denoted as A, B, C, and D, respectively, are involved. All the data of the four vendors were collected from Asian women. The data of vendors A, B and C are set as seen domains whereas the vendor D is treated as unseen domain. To evaluate the generalization capability of our method, we also involve the public dataset, INbreast \cite{moreira2012inbreast}, denoted as E as unseen dataset. In total, this study involves 28,700 mammograms, where 27,000 unannotated images are used for style transfer and contrastive learning. The remaining 1,700 annotated images are adopted for the training/validation/testing of the detection tasks. We conduct a preprocessing step to align various mammograms from A, B, C, D, and E into the same pixel spacing of 0.1 mm to facilitate the training of deep learning methods. For the assessment of detector, we adopt the mean average precision(mAP) metrics for the quantitative comparison.
\begin{table}
\caption{Details of usage for the data involved in this study. Specifically, the columns "Style Transfer" and "Self-Supervision" suggest the number of images involved in the training of CycleGAN and contrastive learning, respectively.}\label{dataset}
\centering
\begin{tabular}{c|c|c|c|c|c}
\hline
Domain & Dataset & Style Transfer & Self-Supervision & Detection (train/val/test) & Vendor \\ \hline
& Style A & 1000 & 8000 & 360/40/100 & GE \\
Seen & Style B & 1000 & 8000 & 360/40/100 & UIH \\
& Style C & 1000 & 8000 & 360/40/100 & HOLOGIC \\
\hline
Unseen & Style D & 0 & 0  & 0/0/100 & SIEMENS \\
& Style E & 0 & 0 &  0/0/100 & INbreast \\
\hline
\end{tabular}
\end{table}

\subsection{Ablation Study}

The ablation experiments are composed of two parts. First, we evaluate efficacy of original simple contrastive learning (SimCLR) \cite{chen2020simple} on the lesion detection tasks. Specifically, we compare lesion detection performance with 1) no pretraining, 2) ImageNet, 3) SimCLR on MammoPre, and 4) SimCLR on $\rm ImageNet\rightarrow MammoPre$ pretrained models and report the corresponding performance in the first to fourth rows of Table~\ref{tab:ablation}. The third row 
suggests that SimCLR is trained from scratch with the unlabeled images of vendors A, B and C, where these unlabeled set is denoted as MammoPre. The fourth row $\rm ImageNet\rightarrow MammoPre$ indicates that SimCLR is initialzed with ImageNet parameters and then trained with the MammoPre set. In this part, the pretrained model of $\rm ImageNet\rightarrow MammoPre$ can yield better detection performance.

The second part is to assess the effectiveness of each component of our framework. Specifically, we compare the pretrained models derived from multi-style contrastive learning (MSCL), multi-view contrastive learning (MVCL) and the combination of multi-style and multi-view contrastive learning (MSVCL) w.r.t. the lesion detection performances, which are listed in the fifth to seventh rows in Table~\ref{tab:ablation}, respectively. As can be found in Table~\ref{tab:ablation}, the effectiveness of our contrastive learning for the lesion detection tasks on both seen and unseen domains is corroborated.
The detailed ablation analysis w.r.t. the mass and clustered calcifications detection tasks can be found in the Tables 1-2 of the supplement.

\begin{table}
\caption{Ablation analysis w.r.t. the pre-training strategy and our methods. The results on mAP of the mass and clustered calcifications detection tasks are reported. The performances w.r.t. the mass and clustered calcifications detection tasks are shown in the Table 1 and Table 2 of the Supplement, respectively.}\label{tab:ablation}
\centering
\begin{adjustbox}{center}
\begin{tabular}{c|c|c|c|c|c|c|c|c}
\hline
\multirow{2}{*}{Method} & \multirow{2}{*}{Pretrain strategy} &\multicolumn{4}{c|}{Seen domain }&\multicolumn{3}{c}{Unseen domain} \\
\cline{3-9} & & Style A & Style B & Style C & Avg. & Style D & Style E & Avg. \\ \hline
Random & {None} & 0.427 & 0.563 & 0.487 & 0.492 & 0.345 & 0.525 & 0.435 \\
Supervised & {ImageNet} & 0.731 & 0.7805 & 0.728 & 0.746 & 0.673 & 0.811 & 0.742 \\
SimCLR & {MammoPre} & 0.740 & 0.772 & 0.742 & 0.751 & 0.665 & 0.794 & 0.729\\
SimCLR & {$\rm ImageNet\rightarrow MammoPre$} & \textbf{0.749} & \textbf{0.774} & \textbf{0.759} & 
\textbf{0.761} & \textbf{0.687} & \textbf{0.819} & \textbf{0.753} \\
\hline
MSCL & {$\rm ImageNet\rightarrow MammoPre$}& 0.768 & 0.780 & 0.775 & 0.774 & 0.703 & 0.845 & 0.774 \\
MVCL & {$\rm ImageNet\rightarrow MammoPre$}& 0.756 & 0.785 & 0.771 & 0.771 & 0.706 & 0.830 & 0.768 \\
MSVCL & {$\rm ImageNet\rightarrow MammoPre$}& \textbf{0.779} & \textbf{0.812} & \textbf{0.784} & \textbf{0.792} & \textbf{0.717} & \textbf{0.862} & \textbf{0.789} \\
\hline
\end{tabular}
\end{adjustbox}
\end{table}

\subsection{Comparison with State-of-the-Art (SOTA) Methods}

To further compare with other domain generalization methods, three SOTA methods, i.e., BigAug \cite{zhang2020generalizing}, Domain Diversification(DD) \cite{kim2019diversify}, and EISNet \cite{wang2020learning} are implemented. The BigAug \cite{zhang2020generalizing} is a conventional data augmentation method, whereas the DD \cite{kim2019diversify} proposed a generative learning method for domain generalization. The EISNet \cite{wang2020learning} is a learning-based method that explores task-specific and domain-invariant features. Our method on the other hand can decouple the downstream tasks intrinsically and provide task- and domain-invariant features. We trained all methods with the multi-view images from various vendors and explored the domain generalization correspondingly.

The comparison results are shown in Table~\ref{tab:comparison}, where the Baseline row suggests the result of SimCLR with $\rm ImageNet\rightarrow MammoPre$ training strategy. As it can be found, our MSVCL can promise the best lesion detection performance on either seen or unseen domains. In particular, our method outperformed the EISNet by $1.8\%$ on seen domains and $2.2\%$ on unseen domains and data. Therefore, our MSVCL can be a new referential method for the pretraining of useful backbone for the downstream mammographic image analysis problems. 
The details performance comparison w.r.t. the mass and clustered calcifications detection tasks can be found in the Tables 3-4 of the supplement.

\begin{figure}[!t]
\centering
\makebox[\textwidth][c]{\includegraphics[width=1.2\textwidth]{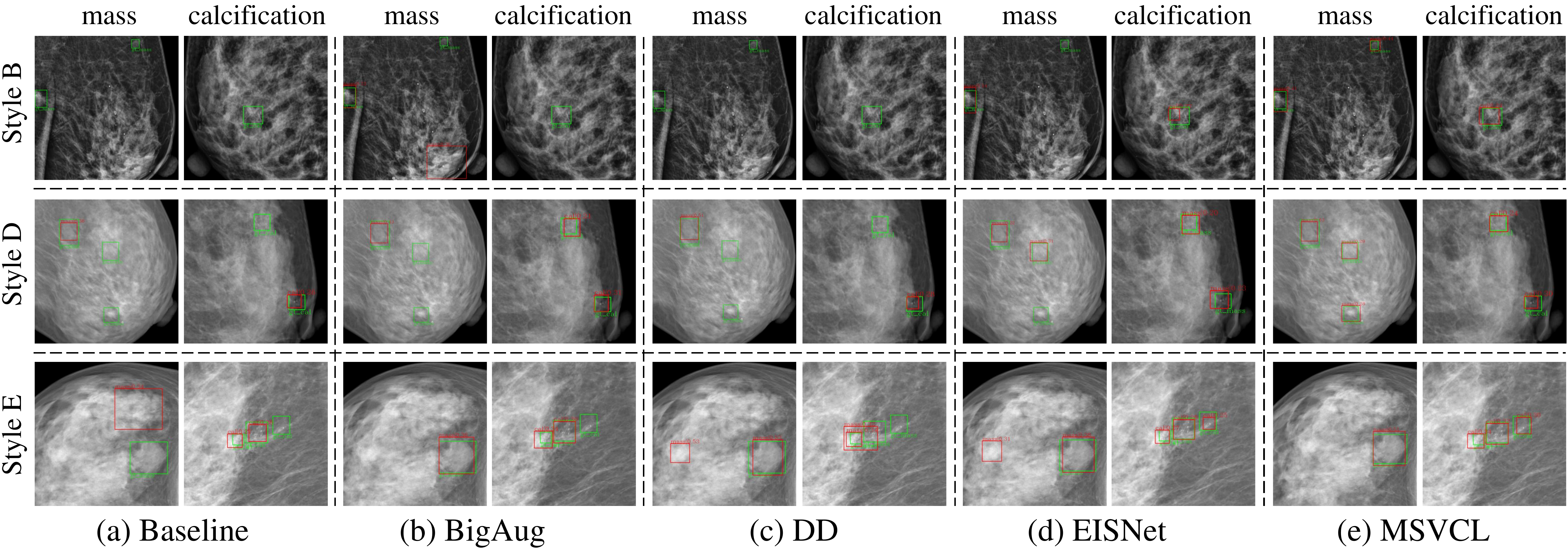}}
\caption{Qualitative results for mass and clustered calcifications detection from different methods on one seen domain (style B) and two unseen domains (style D, E). The green boxes refer to the lesion regions labeled by the radiologist. The red boxes refer to the lesion regions detected by different methods.}
\label{fig:detection_result} 
\end{figure}
 
\begin{table}
\caption{Performance comparison between our MSVCL and other SOTA domain generalization methods. The results on mAP of the mass and clustered calcifications detection tasks are reported. The performances w.r.t. the mass and clustered calcifications detection tasks are shown in the Table 3 and Table 4 of the Supplement, respectively.}\label{tab:comparison}
\centering
\begin{tabular}{c|c|c|c|c|c|c|c}
\hline
\multirow{2}{*}{Method} & \multicolumn{4}{c|}{Seen domain }&\multicolumn{3}{c}{Unseen domain} \\ 
\cline{2-8} & Style A & Style B & Style C & Avg. & Style D & Style E & Avg. \\ \hline
Baseline & 0.749 & 0.774 & 0.759 & 0.761 & 0.687 & 0.819 & 0.753 \\
\hline
BigAug \cite{zhang2020generalizing}  & 0.756 & 0.792 & 0.743 & 0.763 & 0.692 & 0.829 & 0.760 \\
DD \cite{kim2019diversify}  & 0.754 & 0.792 & 0.770 & 0.772 & 0.695 & 0.821 & 0.758 \\
EISNet \cite{wang2020learning}  & 0.765 & 0.781 & 0.775 & 0.774 & 0.690 & 0.845 & 0.767 \\
\hline
MSVCL(ours) & \textbf{0.779} & \textbf{0.812} & \textbf{0.784} & \textbf{0.792} & \textbf{0.717} & \textbf{0.862} & \textbf{0.789} \\
\hline
\end{tabular}
\end{table}

\begin{figure}
  \centering
  \subfigure[Baseline Pretraining]{
    \label{fig:tsne:a} 
    \includegraphics[width=0.3\textwidth]{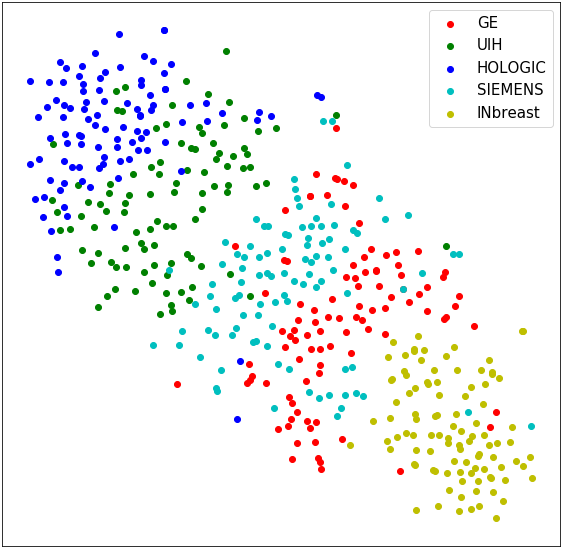}}
  \subfigure[MSVCL Pretraining]{
    \label{fig:tsne:b} 
    \includegraphics[width=0.3\textwidth]{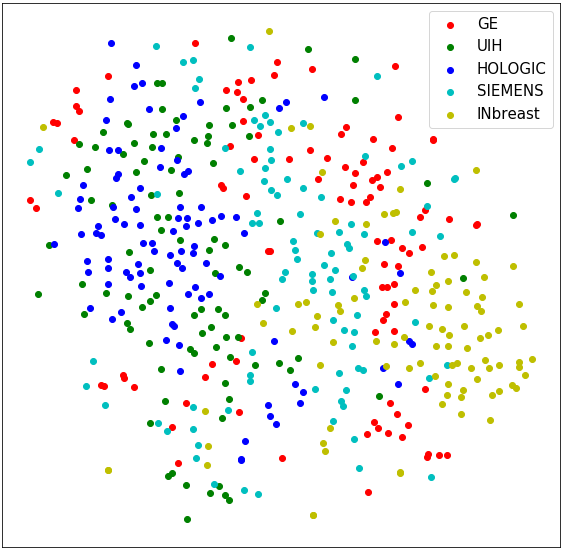}}
  \caption{t-SNE visualization for the pre-trained models from Baseline and MSVCL.}
  \label{fig:tsne} 
\end{figure}

To further visually illustrate the efficacy of our MSVCL pretrained model, the t-SNE \cite{van2008visualizing} is employed to visualized data distribution of various vendor domains in the embedded feature space. Fig. \ref{fig:tsne} compares the pretrained models of Baseline and MSVCL. As can be found, our MSVCL can better break the distribution boundaries between various vendor domains. Therefore, the encoded features shall be more style-invariant. We also demonstrate the detection results visually in the Fig. \ref{fig:detection_result} and the style transfer results of CycleGAN in Fig. 1 of the supplement for more visual assessment.

\section{Conclusion}
A new domain generalization method is proposed to assist the lesion detection schemes in mammography. Specifically, we conduct multi-style and multi-view contrastive learning scheme to embed domain-invariant features to various vendors. The experimental results suggest that our domain generalization method can help to significantly improve the lesion detection tasks on both seen and unseen domains. In particular, for unseen domain of INbreast dataset, we also can achieve the best performance, compared to the three implemented SOTA domain generalization methods. Therefore, the efficacy of our method is corroborated. 

\bibliographystyle{splncs04}
\bibliography{refs}
\end{document}